\documentclass[a4paper,fleqn,usenatbib,useAMS]{mnras}

\usepackage{mathptmx}

\usepackage[T1]{fontenc}
\usepackage{ae,aecompl}

\usepackage{graphicx}	
\usepackage{amsmath}	
\usepackage{amssymb}	

\newcommand{\logg}{\log\,g}
\newcommand{\teff}{T_{\rm eff}}
\newcommand{\bz}{\langle B_z \rangle}

\newcommand{\kms}{km\,s$^{-1}$}

\title[Magnetic Herbig~Ae/Be stars]{
New evidence for weak magnetic fields in Herbig~Ae/Be stars}

\author[S. P. J\"arvinen et al.]{
S.~P.~J\"arvinen$^{1}$\thanks{E-mail: sjarvinen@aip.de},
T.~A.~Carroll$^{1}$,
S.~Hubrig$^{1}$,
I.~Ilyin$^{1}$,
and
M.~Sch\"oller$^{2}$
\\
$^{1}$Leibniz-Institut f\"ur Astrophysik Potsdam (AIP), An der Sternwarte~16, 14482~Potsdam, Germany\\
$^{2}$European Southern Observatory, Karl-Schwarzschild-Str.~2, 85748 Garching, Germany
}

\date{Accepted XXX. Received YYY; in original form ZZZ}

\pubyear{2019}

\begin{document}
\label{firstpage}
\pagerange{\pageref{firstpage}--\pageref{lastpage}}
\maketitle

\begin{abstract}
  In recent years Herbig~Ae/Be stars receive considerable attention as
  their disks are believed to be the sites of on-going planet formation.
  Confirming the presence of magnetic fields in these stars is critical for
  understanding the transport of angular momentum during the protostellar
  phase. Furthermore, magnetic fields set the conditions for strongly
  anisotropic accretion. In this study we present the results of our recent
  observing campaigns of a sample of Herbig~Ae/Be stars aimed at measurements
  of their magnetic fields applying the Singular Value Decomposition method to
  high resolution spectropolarimetric observations. The strongest longitudinal
  magnetic field of 209\,G is detected in the Herbig~Be star HD\,58647,
  whereas the weakest field of 17\,G is measured in the Herbig~Ae star
  HD\,190073. A change of polarity is detected for HD\,58647 and in the
  Herbig~Be star HD\,98922. The obtained results provide further evidence
  that Herbig~Ae/Be stars possess much weaker magnetic fields than their
  lower mass counterpart T\,Tauri stars with magnetic fields of kG order.
  \end{abstract}

\begin{keywords}
  stars: individual: HD\,58647, HD\,98922, HD\,139614, HD\,165133, HD\,190073 --
  stars: magnetic fields --
  stars: pre-main sequence --
  stars: variables: T\,Tauri and Herbig Ae/Be
\end{keywords}



\section{Introduction}
\label{sec:intro}

Herbig~Ae/Be stars are pre-main-sequence objects with pronounced emission line 
features and an infrared excess indicative of dust in their circumstellar disks.
It was previously assumed that these stars are intermediate-mass analogs of
T\,Tauri stars, but with convectively stable envelopes that do not support the
dynamo action found in the fully convective T\,Tauri stars. A number of
magnetic studies have been attempted in the last years, indicating that about
20 Herbig~Ae/Be stars may have globally organized magnetic fields
\citep[e.g.][]{Hubrig2004,Wade2005,Hubrig2009,alecian2013,Hubrig2015,silva2015,silva2018}.
Magnetic fields in these stars might be fossils of the early star formation
epoch, in which the magnetic field of the parental magnetized core was
compressed into the innermost regions of the accretion disks
\citep[e.g.][]{Banarjee}.
Alternatively,
\citet{tout}
proposed a non-solar dynamo that could operate in rapidly rotating A-type
stars based on rotational shear energy.

The compilation of existing magnetic field measurements by
\citet{Hubrig2015}
showed that only very few Herbig~Ae/Be stars have mean longitudinal magnetic
fields stronger than 200\,G, and half of the sample possesses fields of about
100\,G and less. Recent spectropolarimetric observations involving a few
Herbig~Ae/Be stars with relatively low $v\,\sin\,i$ values confirm the results
of this study, revealing the presence of weak, frequently just of the order of
a few tens of Gauss, magnetic fields measured with an uncertainty of only a
few Gauss
\citep{Hubrig2015,silva2015,silva2018}.

In this work, we discuss the results of our recent observing campaigns of a 
sample of Herbig~Ae/Be stars using the High Accuracy Radial velocity Planet
Searcher polarimeter
\citep[HARPS\-pol;][]{snik2008}.


\section{Observations and data reduction }\label{sect:obs}

\begin{table}
\centering
\caption{
  Logbook of the observations. The columns give the name of the star, the 
  heliocentric Julian date (HJD) for the middle of the exposures, the 
  signal-to-noise ratio (S/N) of the spectra, the measured longitudinal
  magnetic field ($\bz$), and the False Alarm Probability (FAP).
}
\label{T:obs}
\begin{tabular}{llc r@{$\pm$}l l}
\noalign{\smallskip}\hline \noalign{\smallskip}
Star & \multicolumn{1}{c}{HJD} & S/N & \multicolumn{2}{c}{$\bz$} & \multicolumn{1}{c}{FAP} \\
        & 2\,400\,000+ & & \multicolumn{2}{c}{(G)} & \\
\noalign{\smallskip}\hline \noalign{\smallskip}
HD\,58647   &  55906.723$^{*}$ & 437 & \multicolumn{2}{c}{---} & \\
                    &  57908.483 & 392 & $-$121 & 11 & $<10^{-5}$ \\
                    &  57910.467 & 466 & 209 & 10 & $<10^{-7}$ \\
                    &  57911.468 & 343 & \multicolumn{2}{c}{---} & \\
\noalign{\smallskip}\hline \noalign{\smallskip}
HD\,98922   &  55706.513$^{*}$ & 379 & $-$33 & 4 & $<10^{-5}$ \\
                    &  57554.588 & 244 & 28 & 7 & $<10^{-6}$ \\
\noalign{\smallskip}\hline \noalign{\smallskip}
HD\,139614  &  57555.722 &  218 & 25 & 3 & $<10^{-8}$ \\
\noalign{\smallskip}\hline \noalign{\smallskip}
HD\,165133  &  57911.813 &  142 & 97 & 9 & $< 10^{-7}$ \\
\noalign{\smallskip}\hline \noalign{\smallskip}
HD\,190073  &  57554.886 & 164 & 23  & 2  & $<10^{-5}$ \\
                     & 57555.848 & 215 & 34  & 2  & $<10^{-10}$ \\
                     & 57908.913 & 367 & 17  & 1  & $<10^{-10}$ \\
                     & 57909.861 & 179 & 20  & 2  & $<10^{-8}$ \\
                     & 57910.919 & 323 & 19  & 1  & $<10^{-10}$ \\
\noalign{\smallskip}\hline \noalign{\smallskip}
\end{tabular}
\begin{minipage}{\columnwidth}
  {\it Note:}
Observations marked with $^{*}$ were already reported in \citet{Hubrig2013}.
\end{minipage}
\end{table}

The new HARPS\-pol observations of the two Herbig~Ae stars HD\,139614 and
HD\,190073, the two late Herbig~Be stars HD\,58647 and HD\,98922, and the
early Herbig~Be star HD\,165133 were obtained on 2016 June 15 and 16, and on
2017 June 3 to 6. Each observation consisted of subexposures with exposure
times varying between about 6 minutes and 47 minutes, depending on the target
visual magnitude. After each subexposure, the quarter-wave retarder plate was
rotated by $90\degr$. The resolving power of HARPS is about $R = 115\,000$,
with spectra covering the spectral range 3780--6910\,\AA{}, with a small gap
between 5259\,\AA{} and 5337\,\AA{}. The reduction and calibration of the
obtained spectra was performed using the HARPS data reduction software
available on La~Silla. The normalization of the spectra to the continuum level
is described in detail by 
\citet{Hubrig2013}.
The summary of the HARPSpol observations is given in Table~\ref{T:obs} in
columns 1 to 3.


\section{Longitudinal magnetic field measurements}
\label{sec:meas}

According to previous studies are the magnetic fields in Herbig~Ae/Be stars
expected to be weak. Therefore, to study the presence of magnetic fields
in these stars, we employed the dedicated Singular Value Decomposition technique
\citep[SVD;][]{Carroll2012}.
The SVD approach is very similar to that of the Principle Component Analysis
(PCA). In this technique, the similarity of the individual Stokes~$V$ profiles
allows one to describe the most coherent and systematic features present in all
spectral line profiles as a projection onto a small number of eigenprofiles.
The excellent potential of the SVD method, especially in the analysis of weak
fields in Herbig~Ae stars, was already presented by
\citet{Hubrig2015}
and
\citet{silva2015,silva2018}.

We have used the Vienna Atomic Line Database
\citep[VALD; e.g.][]{Kupka2011,VALD3}
to construct a line mask for each star based on their stellar parameters.
Furthermore, we have checked that each line in the mask is present in the
stellar spectra. Obvious line blends and lines in telluric regions were
excluded from the line list. The mean longitudinal magnetic field is
determined from the SVD spectra by computing the first-order moment of the
Stokes~$V$ profile according to
\citet[][]{Mathys1989}:

\begin{equation}
\left<B_{\mathrm z}\right> = -2.14 \times 10^{11}\frac{\int \upsilon V
  (\upsilon){\mathrm d}\upsilon }{\lambda_{0}g_{0}c\int
  [I_{c}-I(\upsilon )]{\mathrm d}\upsilon},
\end{equation}

\noindent
where $\upsilon$ is the Doppler velocity in \kms, and $\lambda_{0}$ and$g_{0}$
are the average values for the wavelength (in nm) and the Land\'e factor
obtained from all lines used to compute the SVD profile, respectively. The
results of the magnetic field measurements are presented in Table~\ref{T:obs}
together with the False Alarm Probabilities (FAPs). The FAP is used to
classify the magnetic field measurements
\citep{Donati1992}:
a profile with ${\rm FAP} < 10^{-5}$ is a definite detection, with
$10^{-5} < {\rm FAP} < 10^{-3}$ is a marginal detection, and with
${\rm FAP} > 10^{-3}$ is a non-detection. The older data for HD\,58647 and
HD\,98922 from
\citet{Hubrig2013}
are used for comparison with the new measurements. In the following we discuss
the magnetic field measurements for each star individually.

\subsection{HD\,58647}

HD\,58647 is a late Herbig~Be star exhibiting double-peaked profiles in some
hydrogen emission lines, such as H$\alpha$ and Br$\gamma$
\citep[e.g.][]{Gradyet, Brittain}.
Linear spectropolarimetric observations of H$\alpha$ showed a large
polarization change across the line
\citep{Vink2002, Mottram2007, HarKuhn2009}.
The detected change in the linear polarization was interpreted in terms of a
compact source of line photons that is scattered off a rotating accretion
disc. A first measurement of the longitudinal magnetic field
$\bz =218\pm69$\,G in HD\,58647 using the moment technique
\citep{Mathys1994}
was published by
\citet{Hubrig2013}.
\citet{Kurosawa2016} 
showed that a model with a small magnetosphere and a disc wind with its inner
radius located just outside of the magnetosphere can well reproduce the
observed Br$\gamma$ profile, wavelength-dependent interferometric
visibilities, colour differential phases, and closure phases simultaneously. 

\begin{figure*}
 \centering 
        \includegraphics[width=1.0\textwidth]{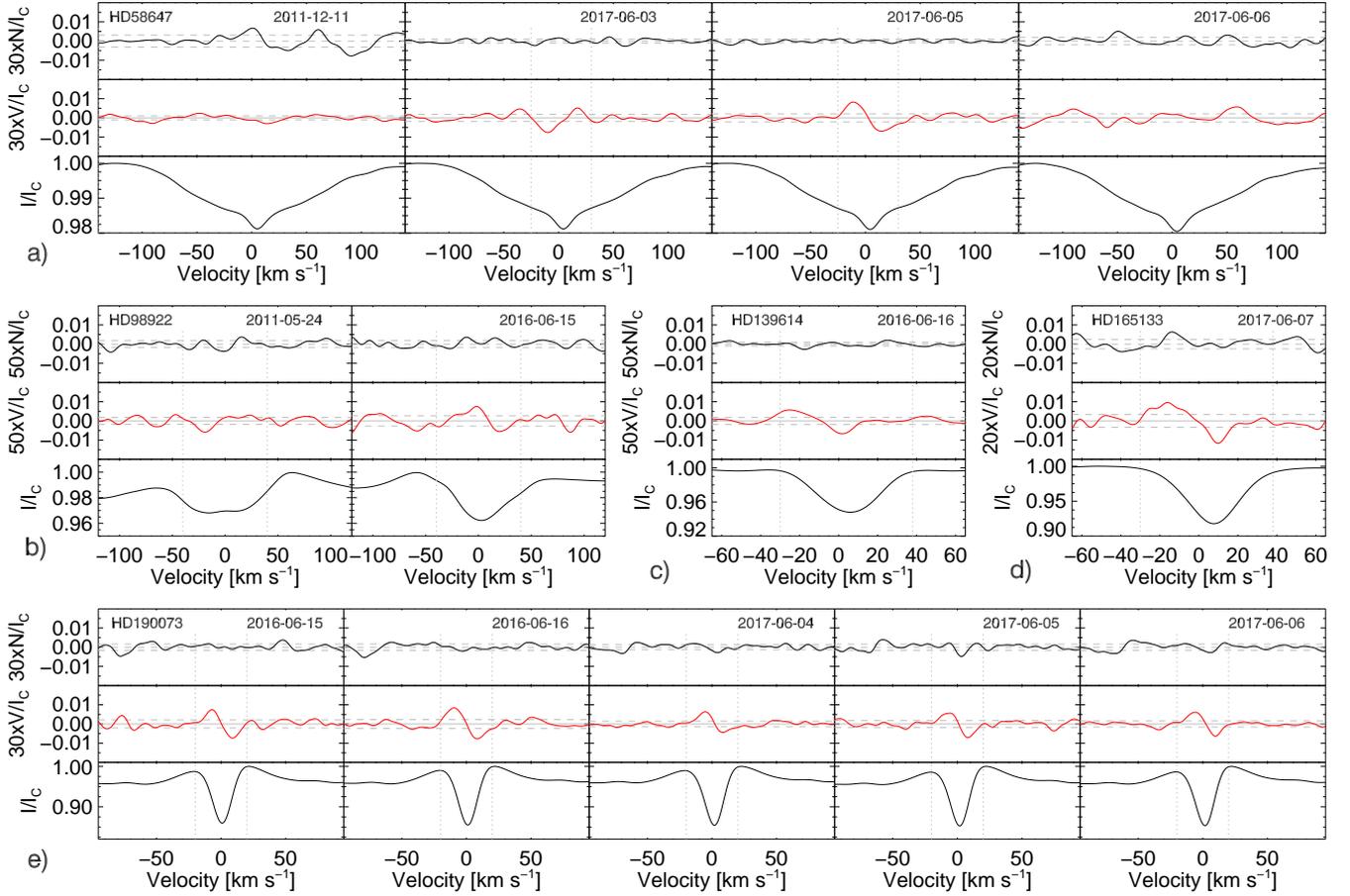}
        \caption{
          SVD Stokes $I$ (bottom), $V$ (middle), and diagnostic null (N)
          profiles (top) for the five Herbig~Ae/Be stars in our sample.
        }
   \label{fig:mos}
\end{figure*}

The line mask is based on $\teff =10500$\,K 
\citep{merin}
and $\logg=3.33$
\citep{Montesinos}. 
The SVD profiles have been calculated using 206 unblended lines with an
average Land\'e factor $\bar{g}_{\rm eff} = 1.12$. As we show in
Fig.~\ref{fig:mos}a, the spectral lines have a rather complex structure with
sharp absorption components in the line cores. Only the observations obtained
on 2017 June~3 and 5 reveal the presence of Zeeman features. The measured
magnetic field strengths show a change of polarity within one day from
$\bz =-121\pm11$\,G to $\bz =209\pm10$\,G. We assume that the sharp cores
observed in the line profiles are originating in the stellar photosphere.

\subsection{HD\,98922}

HD\,98922 is the second late Herbig~Be star in our sample. A first measurement
of the longitudinal magnetic field in this star, $\bz =-131\pm34$\,G, was
published by
\citet{Hubrig2013} 
using the moment technique, whereas 
\citet{alecian2013}
reported a non-detection. As discussed by 
\citet{Hubrig2013}, 
\ion{Fe}{ii} and \ion{Ti}{ii} lines in the high resolution HARPS spectra of
this star show traces of splitting and lines belonging to different elements
exhibit different shapes of their profiles, suggesting that some elements are
inhomogeneously distributed on the stellar surface. The interferometric study
by 
\citet{kraus} 
indicated that the size of the  Br$\gamma$-emitting region is consistent with 
magnetospheric accretion.

Our line mask is based on $\teff \approx 10500$\,K 
\citep{vanden}
and $\logg=4.0$ 
\citep{alecian2013}. 
The SVD profiles have been calculated using 202 unblended lines with an
average Land\'e factor $\bar{g}_{\rm eff} = 1.18$. The analysis of the SVD
profiles (Fig.~\ref{fig:mos}b) reveals a complex and variable shape of the
Stokes~$I$ profiles. The longitudinal magnetic field shows different magnetic
field polarity when comparing the data from 2011 with $\bz =-33\pm4$\,G to
data from 2016 with $\bz =28\pm7$\,G.

\subsection{HD\,139614}

HD\,139614 is a late Herbig~Ae star for which previously published magnetic
field measurements indicated both non-detections and detections
\citep{Wade2005,Hubrig2004,Hubrig2007,Hubrig2009, alecian2013aa}.
\citet{Wade2005} 
failed to detect a magnetic field in HD\,139614 with quoted uncertainties of
25\,G on two consecutive nights, whereas 
\citet{Hubrig2004,Hubrig2007}
reported the presence of a weak magnetic field in the range from $-$116\,G to
$-$450\,G. 

HD\,139614 is probably possessing a transition disk, where the dust gap has
been opened by a single giant planet. Interferometric observations obtained
with the MIDI instrument on the Very Large Telescope Interferometer together
with temperature-gradient modeling favoured a two-component disk structure
with spatially separated inner and outer dust components
\citep{Matter2014}. 
The disk inclination was found to be 20\degr, signifying that we see the target
close to face-on. Later
\citet{Matter2016}
performed a first multi-wavelength modeling of the dust disk and confirmed a gap
structure. The CRyogenic high-resolution InfraRed Echelle Spectrograph
(CRIRES) observations by
\citet{Carmona2017},
made at the European Southern Observatory's Very Large Telescope (VLT),
suggested the presence of an embedded $<2 M_{\rm J}$ planet inside the dust gap
at a very close distance of about 4\,AU.

HD\,139614 is the coolest target among this sample. Our line mask is based on
$\teff =7600$\,K and $\logg =3.9$
\citep{folsom}.
The SVD profiles (Fig.~\ref{fig:mos}c) were calculated using 2870 lines with
an average Land\'e factor $\bar{g}_{\rm eff} = 1.20$. The analysis of the SVD
profiles indicates a definite detection of the longitudinal magnetic field
with $\bz =25\pm3$\,G (FAP $< 10^{-8}$). 

\subsection{HD\,165133}

HD\,165133 is an early Herbig~Be star listed in the catalogue of members
and candidate members of the Herbig~Ae/Be stellar group by 
\citet{Theet}.
Already in the late 1930s it was reported to belong to the Sagittarius region
of the Milky Way
\citep{Sagittarius}. 
The study by 
\citet{Chen}
confirmed its membership in the extremely young open cluster NGC\,6530 at the 
age of 1.5--2\,Myr, which is the dominant cluster in the Sgr\,OB1 association
and is located in the eastern part of the Lagoon nebula. Fig.~\ref{fig:mos}d
shows the detection of a weak longitudinal magnetic field with
$\bz =97\pm9$\,G. Our line mask is based on the B2 spectral type indicated in
the SIMBAD database. The SVD profiles were calculated using 103 spectral lines
with an average Land\'e factor $\bar{g}_{\rm eff} = 1.21$.

\subsection{HD\,190073}

The absorption and emission spectrum of the Herbig~Ae star HD\,190073 was
studied in detail by
\citet{catala2007}
and
\citet{CowleyHubrig}.
This star has a very low projected rotational velocity
($v\,\sin\,i \le 9$\,\kms), which may indicate either a very slow rotation or
a very small inclination of the rotation axis with respect to the line of
sight. The latter option is supported by interferometric observations
\citep{Eisner2004}
as well as by IUE observations
\citep{Hubrig2009}.

A first measurement of the longitudinal magnetic field $\bz =84\pm30$\,G
in HD\,190073 was published by
\citet{Hubrig2006}
and was based on spectra obtained with the low-resolution FOcal Reducer 
low dispersion Spectrograph
\citep[FORS\,1;][]{1998Msngr} 
at the VLT.
\citet{Wade2007}
reported a non-detection, but the following studies by
\citet{catala2007}, \citet{Hubrig2009}, and \citet{alecian2013}
reported definite detections of longitudinal magnetic fields with strengths 
ranging from 111\,G to $-$35\,G.

The observations obtained during 2011 and 2012
\citep{silva2015}
showed a definite detection $\bz =-8\pm6$\,G with a FAP smaller than $10^{-8}$
and a marginal detection $\bz =-15\pm10$\,G with a FAP of $8\times10^{-4}$.
Since the variability time scale of the magnetic field of this star is not yet
known, we decided to monitor it on additional epochs. The line mask applied in
our study is based on $\teff =9250$\,K and $\logg =3.5$
\citep{acke}.
The SVD profiles (Fig.~\ref{fig:mos}e) using new data from 2016 and 2017 are
calculated using 688 lines with an average Land\'e factor
$\bar{g}_{\rm eff} = 1.21$ and show definite detections of positive fields ranging
from $17\pm1$\,G to $34\pm2$\,G.

\subsection{Spectral variability on short and long-term scales}

\begin{figure*}
 \centering 
        \includegraphics[width=1.0\textwidth]{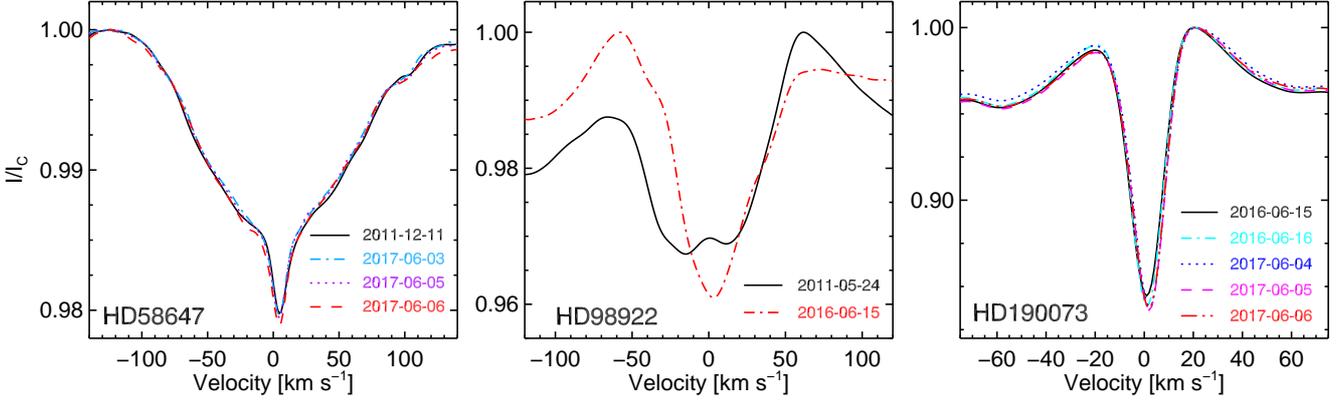}
        \caption{
          Overplotted SVD Stokes $I$ profiles for our sample stars with
          multiple observations.
        }
   \label{fig:Ivar}
\end{figure*}

\begin{figure}
 \centering 
        \includegraphics[width=1.0\columnwidth]{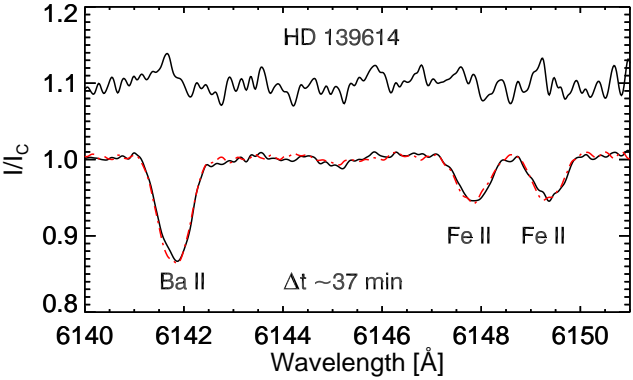}
        \caption{
          {\em Bottom:} Two overplotted subexposures for
          HD\,139614 separated by 37\,min.
          {\em Top:} Difference between the two spectra
          expanded by a factor of 3. The detected differences in the line
          profiles indicate that the star is probably pulsating.
        }
   \label{fig:puls}
\end{figure}

Similar to the spectroscopical behaviour of other Herbig~Ae/Be stars 
\citep[e.g.][]{Hubrig2011a,Hubrig2012,silva2016,silva2018,silva2019},
the Herbig stars with multiple exposures in our sample show variable line
profiles. As we show in Fig.~\ref{fig:Ivar}, the strongest changes in the SVD
Stokes $I$ profiles are detected in HD\,98922. Since a number of Herbig~Ae
stars are known to exhibit $\delta$\,Scuti-like pulsations
\citep[e.g.][]{zwintz},
we also checked the variability on short time scales in the
spectropolarimetric subexposures. Only HD\,139614 shows clear differences in
the observed line profiles recorded in subexposures separated by 37\,min (see
Fig.~\ref{fig:puls}).


\section{Discussion}
\label{sec:disc}

Magnetic fields play a key role in the processes leading to the formation of
stars and planets. Analytical models and MHD numerical simulations of the
evolution of star forming cores show that the magnetic field is critical for
transporting angular momentum during the protostellar phase and sets the
conditions for strongly anisotropic accretion. The presented magnetic field
measurements -- with the strongest longitudinal magnetic field of 209\,G
detected in the Herbig~Be star HD\,58647 and the weakest field of 17\,G
measured for HD\,190073 -- provide further evidence that Herbig~Ae/Be stars
possess much weaker magnetic fields than their lower mass counterpart T\,Tauri
stars, with magnetic fields of kG order.

While for strongly magnetic T\,Tauri stars the model of magnetically driven
accretion and outflows successfully reproduce many observational properties,
the mechanism of mass transfer within the Herbig~Ae/Be star + disk system is
unclear due to the limit of weak or even absent magnetic fields. Yet the small
number of detected magnetic Herbig~Ae/Be stars can be explained not only by
the weakness of their magnetic fields but also by too large measurement errors.
\citet{alecian2013}
reported that about 90\% of their sample of 70 stars do not show detectable
magnetic fields. However, the uncertainty of the magnetic field determinations
in their study was worse than 200\,G for 35\% of the measurements and for 32\%
of the measurements it was between 100 and 200\,G, i.e.\ only 33\% of the
measurements showed a measurement accuracy below 100\,G.
As we have shown in our previous work
\citep{Hubrig2015,silva2015,silva2018},
the uncertainties of the measurements using the SVD technique are dramatically 
better than those achieved with the more commonly used Least Squares
Deconvolution
\citep{Donati1997} 
technique. This can be explained by the way the SVD procedure builds a weighted 
mean SVD line profile. As described by 
\citet{Carroll2012}, 
the resulting mean profiles are calculated by the projection of the observed
line profiles onto the basis of identified signal eigenprofiles 
\citep[Eq.~13 in][]{Carroll2012}. 
The projection coefficients (the inner products) are the weights that each line 
from the line list contributes to the resulting SVD profile. In observed
spectra with varying noise level over the wavelength range it has the effect
that the regions with higher noise have a lower weight according to their
smaller projection coefficients. This is different from the noise weighting in
the least-squares solution of the LSD method.

Furthermore, single snapshot observations are not sufficient to judge whether
a Herbig~Ae/Be star is magnetic or not. The longitudinal magnetic field is 
defined as the disk-integrated magnetic field component along the line of
sight and therefore shows a strong dependence on the viewing angle of the
observer, i.e.\ on the rotation angle of the star. The limitations set by the
strong geometric dependence of the longitudinal magnetic field are usually
overcome by repeating observations several times, so as to sample various
rotation phases, hence various aspects of the magnetic field. Unfortunately,
the rotation periods of Herbig~Ae/Be stars are poorly known
\citep[e.g.][]{Hubrig2011a,Hubrig2011b}. 
The observed light variations in these stars are likely of stochastic nature
and caused by fluctuating disk accretion. Multi-epoch rotation-modulated
longitudinal magnetic field measurements are frequently used to determine
rotation periods, but such monitoring with HARPSpol is possible only in the
framework of a large programme.


\section*{Acknowledgements}

We thank the referee, G.~Mathys, for his valuable comments.
Based on observations made with ESO Telescopes at the La Silla Paranal
Observatory under programme IDs 097.C-0277(A) and 099.C-0081(A).
This work has made use of the VALD database, operated at Uppsala University,
the Institute of Astronomy RAS in Moscow, and the University of Vienna.
This research has made use of the SIMBAD database, operated at CDS,
Strasbourg, France.








\bsp	
\label{lastpage}
\end{document}